\documentclass[preprint]{aastex} 

\usepackage{CJK}
\graphicspath{{Figures/}}
\date{\today}

\begin{document} 
\begin{CJK*}{UTF8}{bsmi}
\title{A Two-Temperature Model of Magnetized Protostellar Outflows}
\author{Liang-Yao Wang (王亮堯)\altaffilmark{1,2}, Hsien Shang (尚賢)\altaffilmark{2}, Ruben Krasnopolsky\altaffilmark{2}, Tzu-Yang Chiang (江子揚)\altaffilmark{2}}

\altaffiltext{1}{Graduate Institute of Astronomy and Astrophysics, National Taiwan University, No. 1, Sec. 4, Roosevelt Road, Taipei 10617, Taiwan; lywang@asiaa.sinica.edu.tw}
\altaffiltext{2}{Academia Sinica, Institute of Astrophysics (ASIAA), and 
	Theoretical Institute for Advanced Research in Astrophysics (TIARA), Academia Sinica, 
	11F of Astronomy-Mathematics Building, AS/NTU. No.1, Sec. 4, Roosevelt Rd, Taipei 10617, Taiwan}

\begin{abstract}

We explore kinematics and morphologies of molecular outflows driven by young
protostars using magnetohydrodynamic simulations in the context of the unified wind
model of Shang et al.  The model explains
the observed high-velocity jet and low-velocity shell features. 
In this work we investigate how these characteristics are affected by the underlying
temperature and magnetic field strength. 
We study the problem of a warm wind running into a cold ambient toroid
by using a tracer field that keeps track of the wind material. While an 
isothermal equation of state is adopted, the effective temperature is 
determined locally based on the wind mass fraction. 
In the unified wind model, the density of the wind is cylindrically stratified and
highly concentrated toward the outflow axis.  
Our simulations show that for a sufficiently magnetized wind, the jet identity
can be well maintained even at high temperatures. However, for a high temperature
wind with low magnetization, the thermal pressure of the wind gas can drive
material away from the axis, making the jet less collimated as it propagates.
We also study the role of the poloidal magnetic field of the toroid. 
It is shown that the wind-ambient interface becomes more resistant to 
corrugation when the poloidal field is present, and the poloidal field that bunches 
up within the toroid prevents the swept-up material from being compressed 
into a thin layer. 
This suggests that the ambient poloidal field may play a role in producing a 
smoother and thicker swept-up shell structure in the molecular outflow.

\end{abstract}


\section{INTRODUCTION}

Bipolar molecular outflows from young stellar objects have been found to
accompany the star formation process since the early, embedded
phase. Understanding their properties may ultimately help shed light on the
central compact protostar-disk systems that is believed to drive the energetic
phenomenon. Historically, wind-blown and jet-driven models have been raised to
account for the different observed morphologies of a wider opening shell or a
collimated jet-like structure \citep{lee2000}. Later, it is found in a young
group of (Class 0) outflows that both features are simultaneous present
\citep[e.g., IRAS 04166$+$2706,][]{santiago2009}. It is therefore important for
models aiming to provide a comprehensive picture of outflow phenomenon to be
capable of explaining such dual characteristics. One model that provides a
natural reconciliation between the jet and wind nature is the unified wind model
proposed by \citet{shang2006} (hereafter paper I). It consists of a wide angle
wind which has a stratified density structure resembling that predicted by the
X-wind theory \citep{shu2000}. The wind is assumed to first run into an ambient
toroid structure expected to form from a dense core that is partially supported
by an ordered magnetic field which it inherited from the parent cloud. The dual
features can be naturally reproduced in this scenario where the
axially concentrated part of the wind forms a collimated jet while the more
tenuous diverging part of the wind impacts upon the ambient and forms a
low-velocity shell. The model has been successfully demonstrated using
magnetohydrodynamic (MHD) simulations in paper I.

Outflow properties can vary from source to source on top of the common
characteristics of bipolarity and jet/shell features. The diversity in cavity
shape, opening angle, kinematic pattern, knot structure, etc.\ can in principal
reflect the underlying physical conditions as well as the evolution history. An
important key to extract information from the rich phenomenon is to build the
connection between physical factors and their consequences on a larger
observable scale. Numerical experiments of outflow models are particularly
valuable in this respect. In paper I it has been demonstrated that the unified
wind model can provide a natural explanation of the jet and shell features, but
the roles of different physical parameters remain to be explored. The magnetic
field is core to the wind model and its role is naturally of great interests. On
the other hand, molecular line studies have suggested a possible warm gas
content in the molecular jet \citep{nisini2007}, which also motivates an
extension of the cold simulations in paper I to a higher temperature
regime. Therefore in this work we examine the consequences of varying the
magnetic field strength and introducing a different gas temperature in context
of the unified wind model of paper I. Meanwhile, the poloidal field of the
toroid may also play a role in the wind-ambient interaction and we also explore
its effects.

We briefly introduce the theoretical background and the numerical setup in
section \ref{sec_theoretical_background} and \ref{sec_numerical_setup}, and
present our results in section \ref{sec_results}. We discuss the implication in
section \ref{sec_discussion} and conclude in section \ref{sec_conclusion}.

\section{MODEL BACKGROUND}\label{sec_theoretical_background}


\subsection{Asymptotic Structure of Magnetocentrifugal Winds}

Magnetocentrifugal wind is probably the most promising mechanism to date to
explain the launching of bipolar outflows from the young stellar objects. As in
paper I, we adopt the same limiting case of the asymptotic behavior predicted by
the X-wind theory. The density, toroidal magnetic field, and velocity of the
wind take the following forms:
\begin{equation} \label{eqn_wind}
  \rho(\varpi)=\frac{D_{0}}{\varpi^{2}}, \quad
  B_{\phi}(\varpi)=\frac{B_{0}}{\varpi}, \quad
  v_{w}=V_{0},
\end{equation}
where $\varpi=r \sin\theta$ is the cylindrical radius from the axis, and the
wind is in radial direction. The density constant $D_0$ is related to and is
determined by the adopted (two-sided) wind mass-loss rate $\dot{M}_W$ through
$\dot{M}_W=2\pi D_0 V_0\ln|(1+\cos\theta_1)/(1-\cos\theta_1)|$. The limiting
angle $\theta_1$ is taken to be half the first cell size of the
$\theta$-grid. The magnetic field strength constant $B_{0}$ is related to the
density constant through $B_{0}=\sqrt{4\pi D_{0}}\,V_{0}/M_{A}$ (in Gaussian
convention), where $M_{A}$ is the Alfv\'{e}n Mach number of the wind. Thus for a
given mass-loss rate (a given density constant), the magnetic field strength is
inversely proportional to $M_A$. We adopt a constant velocity of
50\,km\,s$^{-1}$ for the wind, and with the chosen velocity value we further fix
the wind density constant $D_0$ for a wind mass-loss rate of
$3\times10^{-6}$\,M$_{\sun}$\,yr$^{-1}$. The initial toroidal magnetic field
strengths is $\sim0.75/M_A$ gauss at $\varpi=1$\,AU from the outflow axis given
the adopted values.

\subsection{Singular Isothermal Toroids}

For the young outflows from the embedded system, the wind will first interact with the protostellar envelopes, which are part of the star forming core yet to be accreted onto the central protostar-disk. When dynamically important magnetic fields are present during core formation, anisotropic mass distribution is expected to form due to the easier settling of mass along the field lines than across. This problem was studied by \citet{li1996} and an equilibrium self-similar model was proposed to describe the core density $\rho(r,\theta)$ and magnetic flux function $\Phi(r,\theta)$. The radial and angular dependence of the functions are assumed to take a separable form of
\begin{equation} \label{eqn_toroid}
  \rho(r,\theta)=\frac{a^{2}}{2\pi Gr^{2}}R(\theta),
  \quad
  \Phi=\frac{4{\pi}a^2r}{G^{1/2}}\phi(\theta),
\end{equation}
where $R(\theta)$ and $\phi(\theta)$ are dimensionless angular functions.
The poloidal field is obtained through
$B(r,\theta)=\frac{1}{2\pi}\nabla\times(\frac{\Phi}{r\sin\theta}e_\phi)$.
The possible solutions are characterized by a single parameter $H_0$
representing the fractional overdensity supported by the magnetic field above
that by the thermal pressure. An equivalent parameter $n$ ($n=4H_0$) is
sometimes used in place of $H_0$, and a larger $n$ corresponds to a more
flattened core structure. We adopt the density distribution of $n=2$ toroid as
the initial condition. Simulations are carried out both with and without
including the poloidal magnetic field threading the matter. Such ambient field
was ignored in paper I for simplicity, and we will examine its role here.

\section{SIMULATION SETUP}\label{sec_numerical_setup}

The problem of a warm magnetized wind running into an ambient toroid is studied
using magnetohydrodynamic (MHD) simulations. We evolve the 2D axisymmetric
problem in spherical coordinate ($r,\theta,\phi$) with the outflow axis lying at
$\theta=0\arcdeg$. The usual set of ideal MHD equations are solved using the
ZEUS-TW code \citep{krasnopolsky2010}.  A scaled grid of 1600 points is used to
cover the computation domain from $r=1.5$\,AU to $2\times10^5$\,AU. The finest
grid spacing is 0.08 AU at the innermost cell. We only consider one side of the
bipolar outflow and the $\theta$ grid is in uniform step of 0.125$\arcdeg$
across $\theta=0\arcdeg$ to $90\arcdeg$. The initial condition is set up with a
toroid with zero velocity. The density distribution is always initialized with
the toroid solution while the poloidal field threading it is ignored in the
first part of our exploration. We later include this poloidal field component to gain insight to its
role. The boundary conditions on the outflow axis and at the equator are set by
symmetry. Outflow conditions are set for the outer $r$ boundary. A wind
is imposed at the inner $r$ boundary by setting the density and toroidal
magnetic field values with the wind solutions in every time step. The inner $r$
boundary conditions of the ${\bf V}\times{\bf B}$ are set by continuity to avoid anchoring
the poloidal filed lines.

There are several differences in terms of numerical setup between this work and
paper I. For example, we have moved from the ZEUS-2D code \citep{stone1992} to
ZEUS-TW in order to take advantage of the much faster production speed of
parallel computing. This allows us to finish high resolution tests within
affordable time. We have also adopted a spherical coordinate system instead of a
cylindrical one to accommodate the radial wind velocity more easily. The wind
zone, a box that was used to adapt the spherical wind into a cylindrical grid,
is abandoned, and the winds are imposed through boundary conditions (the ghost
zones). We checked that the new numerical implementation produces nearly
identical output for the same problem setup. Hence the results in this work should
be consistent with that of paper I.

For our study of a warm gas propagating into a cold medium, two characteristic
temperatures are specified. For the ambient toroid we adopt a value of
$a_{\rm ambient}=2\times10^{4}$\,cm\,s$^{-1}$ ($\sim10$\,K), which is also the
global isothermal sound speed used in paper I. We adapt the code by introducing a
second sound speed $a_{\rm wind}$ to evolve the wind while retaining an
isothermal equation of state. This is done by allowing an effective temperature
to be determined locally based on the presence of wind or ambient material at
each position. An evolving tracer field ($\rho_{\rm wind}$) is used to keep track of the wind material and
facilitate this prescription. The gas is cold if it comes from the ambient
toroid and warm if injected as a wind. With these considerations, we come to
adopt the formula of
  $$ a=\sqrt{f a_{\rm wind}^{2}+(1-f)\,a_{\rm ambient}^{2}} $$
for the effective sound speed. The parameter $f$ is the wind-originated mass
fraction defined as $f\equiv\rho_{\rm wind}/\rho$. 
Such formula allows the pure toroid and wind material to
evolve under their own isothermal sound speed, while a simple linear
interpolation between the two temperatures is used for region of mixing. Note
the simulations essentially reduce to a globally isothermal condition when
$a_{\rm wind}$ is set equal to $a_{\rm ambient}$. As mentioned, to enable
calculation of this two-temperature isothermal scheme, we keep track of the wind
material by using a tracer field. We initialize the tracer to zero at $t=0$
in the simulations when no wind material is yet injected. The same wind boundary condition
used to set the density at the inner radius is also applied to the tracer field
in every time step. Since it simply follows the matter transport passively, it
keeps a copy of all wind-originated material.

\section{NUMERICAL RESULTS}\label{sec_results}

In the following subsections we present our simulation results. First we
demonstrate the general features of the warm magnetized wind simulation with one
example case. We then explore the consequences of varying the wind temperature
and toroidal magnetic field strength of the wind by examining a group of runs
with different $a_{\rm wind}$ and $M_{A}$. 
Probing the relative importance of thermal and magnetic pressures in each cases 
using for instance the plasma beta parameter ($\beta_{\rm plasma} \equiv8\pi a^2\rho/B^2$) may help 
provide insights.
Note in the first part of our explorations (sections \ref{sec_results_basic} 
and \ref{sec_results_mcs1}) the poloidal magnetic field threading the ambient 
medium is not included. In section \ref{sec_results_bp} we initialize the
poloidal magnetic field with the proper toroid solution of \citet{li1996}.

\subsection{Basic Features}\label{sec_results_basic}

\begin{figure}     
  \includegraphics[width=\textwidth]{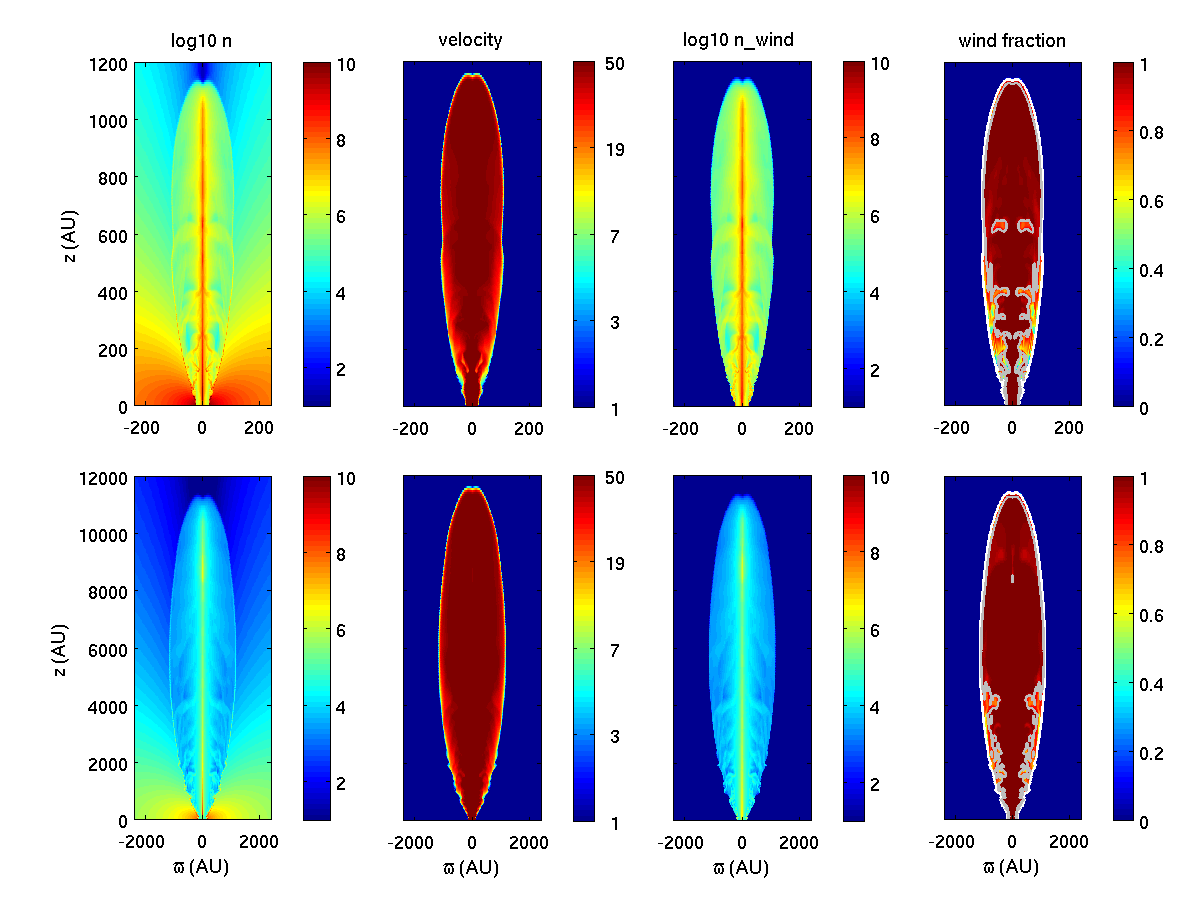}	
  \caption{Maps of (\textit{left to right}) density (cm$^{-3}$), 
  	poloidal velocity magnitude (km\,s$^{-1}$), 
  	tracer field (wind density $\rho_{\rm wind}$) (cm$^{-3}$), and wind fraction 
  	for the reference case of $M_A=30$ and
    $a_{wind}=1.2\times10^5$\,cm\,s$^{-1}$ (T$\sim400$\,K). The upper
    and lower rows are snapshots at $t=100$ and $t=1000$\,yr,
    respectively. The white and gray lines overlaid on the wind
    fraction maps are 10\% and 90\% contours.}
  \label{fig_onecase}
\end{figure}

As a demonstration of the general results, we present snapshots of a typical
simulation at $t=100$ and 1000\,yr in the upper and lower row panels of
Figure \ref{fig_onecase}, respectively. For this particular case the wind sound
speed is $a_{\rm wind}=1.2\times10^{5}$\,cm\,s$^{-1}$ ($\sim400$\,K) and the
Alfv\'{e}n Mach number is $M_{A}=30$. From left to right columns are maps of the
density, magnitude of poloidal velocity (in km\,s$^{-1}$), the tracer field 
$\rho_{\rm wind}$, and the wind fraction $f$. The density and wind density maps 
in this work are plotted as number density of molecules assuming a mean 
molecular weight of $3.88\times10^{-24}$\,g.

The basic characteristic of jet and shell features can be clearly
identified in the density map. The collimated structure concentrated on the
$z$-axis is a result of the intrinsic wind density profile (equation \ref{eqn_wind}). Although the initial
velocity is in radial direction, one does not see a significant density drop
(near the axis) expected for a spherically expanding wind. This is due to
magnetic collimation of the toroidal field in the wind which helps maintain this
cylindrically stratified density structure. 
Moving away from the central axis, the flow density decreases with
cylindrical radius until reaching a higher density structure at the
wind-ambient interface, which is the swept-up shell. 
If we further compare the upper and lower
panels, one can see the clear resemblance between the swept-up shell at
different epochs of $t=100$ and 1000\,yr. As pointed out in paper I, this
self-similar behavior is not surprising since both the asymptotic wind and the
toroid solutions are scale-free. The wind propagates mostly near its initial
speed of 50\,km\,s$^{-1}$ until dropping abruptly at the wind-toroid interface,
as shown by the velocity map. This confirms that the wind-ambient interaction
still lies within the supersonic/superalfv\'{e}nic regime in our new simulation
as in the cold flow of paper I. This example case basically demonstrates that
the overall dynamic of the unified wind model is qualitatively unchanged when we
increase the wind temperature from $\sim$10 K (in paper I) to $\sim$400 K.

In the third column of Figure \ref{fig_onecase} are snapshots of the tracer
field. This field is a critical component in our simulations which facilitates
the calculation of effective temperatures and sound speeds in our
two-temperature prescription. Moreover, it keeps a record of the distribution of
wind-originated material which can not be seen in the density maps. The figures
show that the primary wind is confined within a limited volume. The values of
the tracer is essentially zero or negligible outside this region, but appear to
be very similar or identical to the density maps (in the first column) within
the region. Note while the central jet component is clearly seen in the tracer
field maps, the shell feature is not. This implies different compositions and
origins of the two characteristic features. While the jet structure is
essentially part of the primary wind, the shell consists mostly of the swept-up
ambient material which does not show up in the tracer field maps. The boundary
where the density of wind material drops to near zero is close to the inner edge
of the swept-up shell.

Another way to visualize the wind distribution is to examine the relative mass
composition at each position. We show the results in
the fourth column of Figure \ref{fig_onecase}. 
The wind fraction of 10\% and 90\% are highlighted with white and gray contours. 
It is clear from the resulting maps that the majority part of the simulation are either
wind-dominated (within the gray contour, $f>90\%$) or ambient-dominated
(outside the white contour, $f<10\%$). The transition from wind to toroid usually
happens rather quickly with only little intermediate region. Hence we may
understand the two-temperature simulations as follow: the major part of the wind
evolves isothermally under the higher temperature specified for the wind, while
the toroid retains its original temperature of $\sim10$\,K. Materials in between
the two contours are subjected to notable degree of mixing, and are the region
that will take on the interpolated temperature value in the simulation.

While the division between wind and ambient appears obvious, there are structures
with notable degree of mixing located within the wind-dominated region. For example, at
a height of $z\simeq350, 450,$ and 700\,AU in the upper panel of wind fraction
map, there are blobs of gas with $f\simeq0.8$. The
velocity map at the corresponding positions shows that these patches have
slower velocities compared to the wind surrounding them. Since they do not
propagate as fast as the major part of the wind, they should lag behind
the main portion of the flow. This is indeed seen in
the lower panel of the wind fraction map which is a snapshot at a later time
frame.
The development of these mixing structures shows that 
the ambient material can be entrained at the strongly sheared wind-ambient interface 
where Kelvin-Helmholtz instability is active. 
Occasionally, larger blobs of ambient materials can also get incorporated into the wind 
and subsequently get ejected along the flow. 
Given the much higher density of the inner toroid than the wind, the entrained 
ambient mass can dominate the local mass fraction for some time 
before being diluted by the wind. 
We discuss this in more detail in section \ref{sec_discussion_mixing}.

\subsection{The Joint Effect of Wind Temperature And B$_\phi$}\label{sec_results_mcs1}

A group of simulations has been carried out to investigate effects of the wind
temperature and the wind magnetic field on the outflow morphology and
dynamics. We present the resulting density snapshots at $t=1000$\,yr in
Figure \ref{fig_mcs1nobp}. Within this three-by-three mosaic, the wind
temperature increases from $\sim$10\,K in column 1 to $\sim$400\,K and
$\sim$2000\,K in column 2 and 3, corresponding to $a_{\rm wind}=2\times10^4,
1.2\times10^5$, and $2.67\times10^5$\,cm\,s$^{-1}$, respectively. Note that the
sound speed of the toroid is $2\times10^4$\,cm\,s$^{-1}$, so the three cases in
column 1 are in fact globally isothermal. The strength of the toroidal magnetic
field in the wind, on the other hand, decreases form $M_{A}=6$ on the top row
({\it a}) to $M_{A}=30$ and 90 in the second and the third row ({\it b} and
{\it c}). The collection of $a_{\rm wind}$---$M_A$ combinations are meant to explore
a wide range of physical conditions. With a high field strength and a low
temperature, the upper-left case {\it a1} represents the magnetic-dominated
situation with negligible gas pressure. The lower-right panel {\it c3} shows the
contrary case where thermal energy dominates. For a temperature of $\sim400$\,K
and $M_A=30$ in case {\it b2} ($\beta_{\rm plasma}\simeq1$), both magnetic field 
and thermal pressure play a role. Case {\it b2} is also the demonstrative case
discussed in the previous section. We directly list the value of $a_{\rm wind}$ and
$M_A$ above each panel for easy reference.

How raising the temperature can affect the wind is best illustrated with the
three cases of relatively weak magnetic fields, namely, those in the third row
of Figure \ref{fig_mcs1nobp}. When the wind temperature increases from $\sim$10\,K ({\it c1}) 
to $\sim$2000\,K ({\it c3}), the change in the jet component is
evident. The initial density distribution of the wind is cylindrically
stratified and takes the form of $\rho\propto\varpi^{-2}$ (see equation \ref{eqn_wind}), 
which is strongly peaked toward the outflows axis. In case {\it c1}
this jet-like structure is clearly seen and is well maintained along the
flow. At higher temperatures, however, we begin to see how the strongly peaked
density profile gradually flattens with distance. In {\it c2} the jet component
becomes rather flat after reaching $z\sim4000$\,AU, and in {\it c3} it flattens
so rapidly that beyond $z\sim2000$\,AU the jet identity is barely
recognizable. The rapid drop of the axial density peak in {\it c2} and {\it c3} is
a result of thermal pressure gradient force which pushes the gas from high
density region toward the lower ones. The strength of the force scales linearly
with temperature and increases over two order of magnitude from 10\,K to
2000\,K, resulting in the strongly modified jet density profile. Another notable
change is that the opening angle of the shell component becomes larger as
temperature increases. As the figure shows, the lateral size of the swept-up
shell is about twice as large in {\it c3} than in {\it c1}.
  
On the other hand, the effects of varying toroidal magnetic field strength in
the wind are most evident when we compare the simulations at a same high wind
temperature of $T\sim2000$\,K, namely, the three cases in the third column of
Figure \ref{fig_mcs1nobp}. The weakly magnetized case {\it c3} presents a jet
component that flattens rapidly with distance due to the strong pressure
gradient force. In case {\it b3} where the wind is more
strongly magnetized, we tentatively see that the jet is maintained to a larger
distance, although the map is complicated by some substructures. In the
most strongly magnetized case {\it a3} where the toroidal field is more than one order of
magnitude stronger than in {\it c3}, the initial jet density structure turns out
to be well-maintained despite the high wind temperature of $\sim2000$\,K. Since
the three simulations start from the same initial wind density profile and evolve
under the same temperature, the pressure gradient force responsible for
smoothing the jet in {\it c3} must also be active in {\it a3}. It is only
because the stronger toroidal magnetic field in {\it a3} has countered the
effect of thermal pressure that the stratified density profile of the jet is
maintained. This same collimation effect has been shown in paper I for the cold
flow, and for a higher wind temperature the effect is even more dramatic.

We have shown that the temperature and magnetic field each plays a role in
determining the fate of jet-like density structure. For example, in contrast to
the clear difference between {\it c1} and {\it c3} due to the increased
temperature, the counterpart pairs of {\it a1} and {\it a3} appear unaffected by
the temperature because of their stronger toroidal field. The joint effect of
the two factors can be better understood with an overview of all nine cases in
Figure \ref{fig_mcs1nobp}. While the temperature varies column by column and the
toroidal field strength row by row, a general trend is found in direction of upper-left to lower-right.
As we move from a magnetic-dominated regime to a thermal-dominated one, 
the peak density of the jet-like structure drops more rapidly with distance. 
To compare the different degrees of flattening, in the upper panel
of Figure \ref{fig_comparedensitydrop} we present on a log-log scale the density profiles
of all nine cases cut perpendicular to the jet at a height of $z=200$ AU. The
three profiles of the most strongly magnetized cases ($M_A=6$, solid lines) are
very similar despite the different temperatures. Case {\it a1} is almost
indistinguishable from {\it a2}, and {\it a3} only shows a small discrepancy
from the other two. They all preserve the two order of magnitude drop in density
from $\varpi=1$ to 10\,AU obtained from the initial wind solution where
$\rho\propto\varpi^{-2}$. For the less magnetized cases of $M_A=30$ (dashed
lines) and $M_A=90$ (dotted lines), the flattened density profiles are clearly
seen for the 400\,K and 2000\,K cases, but not for 10\,K. In fact, all
three cases at 10\,K ({\it a1}, {\it b1}, and {\it c1}) as well as the strongly
magnetized cases ({\it a2} and {\it a3}) all show similar density profiles in
this figure. We also note that a less magnetized wind with a modest 
temperature of $\sim400$\,K (the intermediate case {\it b2}) shows only a mild density drop.  
Although the density profile is modified by the thermal pressure, the change is slow enough
that its jet identity is still clearly seen up to a distance of at least 10000\,AU from the source
(see Figure \ref{fig_onecase}). The jet identity is strongly modified only in case
of a poorly magnetized wind at high temperature.

The jet structure is clearly related to both the
temperature and the toroidal magnetic field strength.
How the relative importance of the two factors is related to the density drop
is examined in the lower panel of Figure \ref{fig_comparedensitydrop}.
We use the wind density at $\varpi\simeq$1 AU in each profile as a proxy to
the degree of jet flattening and arrange the results using their $\beta_{\rm plasma}$ values. 
Different symbols are used for different magnetic field strengths, while different 
wind temperatures are represented by different colors. Beside the nine cases 
that have been discussed, we also include the 
results of non-magnetized wind at the three temperatures explored (the cross 
symbol on the right boundary of the plot). The three extra points
can be thought of as asymptotic results one would obtain
when the magnetic field strength is decreased indefinitely for a fixed wind temperature.   
The plot shows that for magnetic-dominated cases the 
axially concentrated profiles are well maintained and the density values 
are about the same ({\it a1, b1, c1, a2}, and {\it a3}). In the intermediate regime 
the drop in jet density becomes apparent, and the value gradually approaches what
one would obtain in the hydrodynamic limit as thermal pressure 
becomes more and more dominant. In absence of the toroidal field, the jet follows a 
simply trend where it flattens more significantly for a higher wind temperature, as shown 
by the values of the three cross symbols. For different temperatures, the toroidal magnetic 
field strengths sufficiently high to collimate the flow are different. But once the field 
is strong enough the jet density can be kept at a similar high value. 
Although the scattering of the data points shows that jet structure is not simply 
determined by the relative strength of the thermal and magnetic energy in the wind, 
with this ratio we are able to show the opposing roles of the two.

The opposing roles of the magnetic and thermal pressure on the stratified
density structure is closely related to the properties of the wind solution. The
density distribution of the wind takes the form of $\rho(\varpi)=D_0/\varpi^2$,
which naturally produces a pressure gradient force pointing in the positive
$\varpi$ direction. Since its strength is proportional to both the gas
temperature and the density gradient, it is strongest near the axis where the
gradient is high and is more significant at higher temperatures. On the other
hand, with the initial configuration of $B_{\phi}(\varpi)\propto\varpi^{-1}$,
the toroidal field exerts no Lorentz force. But when the thermal pressure pushes
the centrally concentrated gas outward, it also modifies the toroidal field distribution 
and as a consequence a magnetic force can arise. In a strongly magnetized case
a slight modification in the field configuration can result in a strong magnetic
force, making it more capable of countering the flattening of density peak
caused by the thermal pressure.

\subsection{The Role of Ambient Poloidal Fields}\label{sec_results_bp}

The ordered poloidal magnetic field assumed to support and shape the toroid in
the model of \citet{li1996} has been ignored in the simulations so far. As in
paper I, no poloidal field is set up and only the wind-originated toroidal field
is evolved for simplicity. It has, however, been anticipated that the
presence of a poloidal field in the ambient could help resist lateral expansion of
the wind and therefore aid in the shell collimation. In this section we present
simulation results conducted with the poloidal magnetic fields initialized to the
appropriate toroid solution. Both toroidal and poloidal fields are evolved.

In Figure \ref{fig_mcs1bp} we present the density snapshots at $t=1000$\,yr
of a series of simulations with ambient poloidal fields. The adopted combinations
of wind temperatures and toroidal field strengths are the same as those in
Figure \ref{fig_mcs1nobp} so that for each case in Figure \ref{fig_mcs1bp} 
there is a counterpart in Figure \ref{fig_mcs1nobp}. 
A comparison between the two sets of simulations thus straightforwardly shows 
the effects of the poloidal field. 
The most evident change is clearly seen in the swept-up
shell: a much smoother and thicker structure is formed in place of the original
corrugated interface. Such effect is quite general since it occurs in all nine
cases from magnetically to thermally dominated and does not depend on the wind
properties. On the other hand, we find that the ambient field has little or
negligible effects on the evolution of axial density concentration of the
jets. This is expected since the central jet component propagates almost freely
within the outflow cavity and knows little about the ambient. The shell, on the
contrary, is itself a result of wind-ambient interaction and is directly
affected by the ambient poloidal field.

We examine the outflow structure in more detail again using case {\it b2} as an
example. In Figure \ref{fig_bpandshell} we compare maps of density and poloidal
velocity produced either with ambient field ignored (\textit{upper row}) or
included (\textit{lower row}). The poloidal field lines are shown with thin
white contours in the lower panels. To illustrate the region within which the
wind is confined, we overlay the contours of 1\% wind fraction with thick solid
lines (dark in the left panels and red in the right). Beyond this line the wind
have negligible mass fraction, and we take this as a proxy of the wind
boundaries. By comparing the upper and lower density maps we find only tiny
differences in the enclosed regions. At a given outflow height $z$, the
cylindrical radius $\varpi$ of the boundary is marginally larger when the
ambient poloidal field is absent. Although this appears compatible with the
expectation that an ambient field can contribute to the lateral confinement of
the wind, the effect is not significant for the provided poloidal field
strength.

The modified shell structure in the presence of an ambient poloidal field can
also be examined more clearly in Figure \ref{fig_bpandshell}. In the upper
density map we can easily recognize the thin, high-density layer lying just
outside the wind boundary (dark solid line) as the swept-up shell. It consists
mainly of piled up ambient material that has received momentum from the wind. In
the lower panels, however, a smooth parabolic structure is formed instead of a
corrugated thin boundary. This shell-like density structure also consists mainly
of ambient material, but it is now thicker and appears to be threaded by the
poloidal magnetic field (white thin lines). Therefore, by contrasting the two
different sets of simulations, we find that the poloidal field can
help prevent corrugation and results in a smoother interface. This agrees
with the expectation that the presence of a magnetic field along a 
sheared flow can suppress the growth of instabilities and results in
a more organized flow behavior
\citep[see e.g.,][and references therein]{chandrasekhar1961,frank1996}.
In addition, when the
ambient field lines bunch up in the toroids as the wind pushes them aside, the
enhanced magnetic pressure tends to resist compression and prevents the
density layer from being all squeezed together. The field thus acts like a
cushion within the toroid.
 
Finally, we examine the magnitude of poloidal velocity in the right column of
Figure \ref{fig_bpandshell}. The red dashed lines are contours of
$|v_{\rm poloidal}|= a_{\rm ambient}$, and the flow directions are shown with yellow
arrows for positions with velocity magnitude higher than this value. The same
contours are overlaid on the density maps with dark dashed lines as well. It
turns out that this contour well delineate the boundary between the shell
structure and the rest of the toroid. The reason could be that material not yet
affected by the wind would remain nearly static in our simulations, while that
already affected can have a finite speed. In the lower right panel, one finds
that the velocity arrows in the shell point roughly in a direction perpendicular
to the poloidal fields.
This is consistent with the idea that the gas in the expanding shell is accelerated 
by the magnetic force. 
In summary, although the
poloidal field threading the toroid has only limited effect in terms of
confining the wind, it plays a significant role in wind-ambient interaction by
modifying the resulting shell structure. This plot also shows more clearly that
the strong shear happens near the wind boundary (thick solid lines) at the inner
edge of the compressed shell.

\section{DISCUSSION}\label{sec_discussion}

\subsection{Implications on The Jet And Shell}

Accumulating observations have revealed details in many molecular
outflows. While bearing the same basic characteristics of jet and shell, the
variety of cavity shapes, collimation degree, velocity patterns, etc.\ suggests
some underlying differences among different sources. The rich morphologies and
kinematics could be potential probes to the underlying physical conditions or
evolution histories that are not well understood. Nevertheless, we need to
understand the possible influences of different physical factors in order to
learn from these clues. In view of its success in reproducing the basic jet and
shell characteristics, we adopt the framework of the unified wind model to
explore possible signatures of temperatures and magnetic fields. Several
findings from the series of numerical experiments presented in the previous
section may help illuminate the direction.

The central portion of the cylindrically stratified wind in the model is identified 
with the extremely high velocity jet component \citep{bachiller1996} observed in young molecular outflows . 
Our simulations suggest that this density profiles can be modified by
thermal pressure when the temperature is high and the toroidal magnetic field is
not too strong. In the most extreme case such as {\it c3} in Figure
\ref{fig_mcs1nobp}, the jet feature flattens very quickly. It is difficult to
identify a collimated jet beyond a few thousand AU from the origin.
Therefore, the presence of a clear collimated jet would suggest
either the flow is cold enough that pressure is negligible, or the toroidal field 
is strong enough to suppress the effect of gas pressure if the wind is hot. 
It may be worth noting that since a direct detection of
the magnetic field in the wind is difficult, a highly collimated jet-like
structure found to persist at a high temperature could be a clue of the presence
of dynamically important toroidal field. More quantitative prediction would rely
on comparisons between model and observations, which is however expected to be
complicated by the knotty nature of the real sources.

The simulations also pose some interesting questions on the shell structure. In
the shell model of \citet{shu1991}, molecular outflows are the ambient material
driven into expansion by the momentum of a radial wind. The resulting morphology
of the shell is determined by a bipolarity function which considers the
directional input of the wind and the anisotropic distribution of the
ambient. Paper I has numerically demonstrated the formation of such swept-up
shells, and the results appear to be in good agreement with the low velocity
component observed in young outflows such as HH 211 \citep{gueth1999}, 
L1448C \citep{hirano2010}, IRAS 04166$+$2706 \citep{santiago2009,wang2014}
etc. With much improved spatial resolution over paper I, the new simulations can
resolve instabilities on a smaller scale. While the overall size and extent of
the outflow are not affected by the resolutions, we find corrugated shells in
many cases as a result of the finer grids as shown in Figure
\ref{fig_mcs1nobp}. Since the real world wind and toroid density distributions
are unlikely as smooth as the analytic description adopted in the model,
structures of the swept-up shell can in principal be more irregular than what is
shown in the current simulations. This, however, is based on the results
obtained before including the ambient poloidal field in the simulations. In
Figure \ref{fig_mcs1bp}, it is clearly shown that the presence of such field
component can provide support to the boundary and maintain a much smoother
interface between the wind and the ambient. It then appears that a smooth
boundary of the low velocity shell could actually suggest the presence of the
ambient poloidal field, since otherwise a more corrugated structure may be
expected.

\begin{figure}     
  \includegraphics[width=\textwidth]{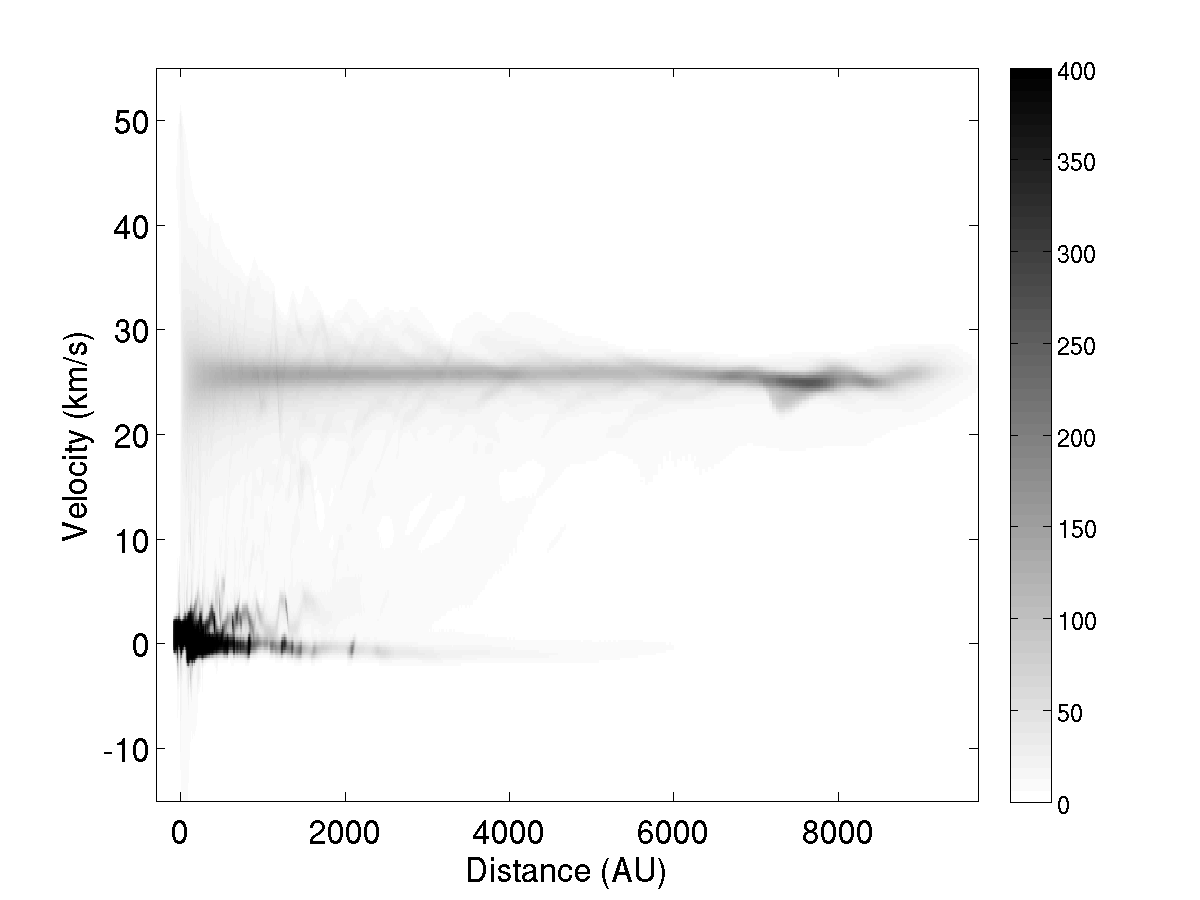}  
  \caption{A synthetic position-velocity diagram cut along the outflow 
	axis of case b2 (without poloidal field). The calculation of 
	CO J $=$ 3--2 line emission assumes a gas temperature of 400\,K 
	and local thermal equilibrium. With an inclination angle of 30{\arcdeg} 
	from the plane of the sky, the dense axially concentrated jet material
	mainly has a projected velocity of $\sim$25\,km\,s$^{-1}$. On 
	the other hand, the swept-up material at low velocity forms 
	another feature near the system velocity which is taken to be zero.}
  \label{fig_PVCut}
\end{figure}

To tell whether these implications may be probed by observations would require 
further predictions of observational features from the model. For example, 
synthetic position-velocity diagrams could be used to interpret the kinematics in 
a more observational fashion. As an example, in Figure \ref{fig_PVCut} we show the 
CO J $=$ 3--2 position-velocity diagram cut along the outflow axis of case b2 (without 
poloidal field). A gas temperature of 400 K and a condition of local thermal 
equilibrium is assumed while calculating the emission. At an 
inclination angle of 30{\arcdeg} from the plane of the sky, the axially concentrated 
jet-like structure shows up with a projected radial velocity of 
$\sim$25\,km\,s$^{-1}$ in the plot. A wider velocity distribution is seen near the base 
(the origin), which is a characteristic of a wide angle wind. On the other hand,
the swept-up material that constitutes the shell has a much lower velocity 
and its emission is largely found near the system velocity which is taken to be zero. 
How the different physical conditions may introduce observable signatures would 
need to be examined more carefully and will be addressed in an upcoming paper.

\subsection{Instability And Mixing Between Wind And Toroid}\label{sec_discussion_mixing}

One question about molecular outflows is whether they are a part of the primary
wind or are made of entrained ambient material. 
With the tracer field, we can examine the wind fraction at each 
position to address this problem. We have seen in Figure \ref{fig_onecase} 
that the major portion of the flow (including the central jet-like structure) 
consists of wind-originated material. On the other hand, 
the shell structure formed at the wind-ambient interface 
has been shown to consist mainly of swept-up ambient material. 
Therefore in the context of the unified wind model, the simple answer to 
the question would be that a molecular outflow is a primary-wind-originated jet 
surrounded by a swept-up shell of ambient material.

\begin{figure}     
  \includegraphics[width=\textwidth]{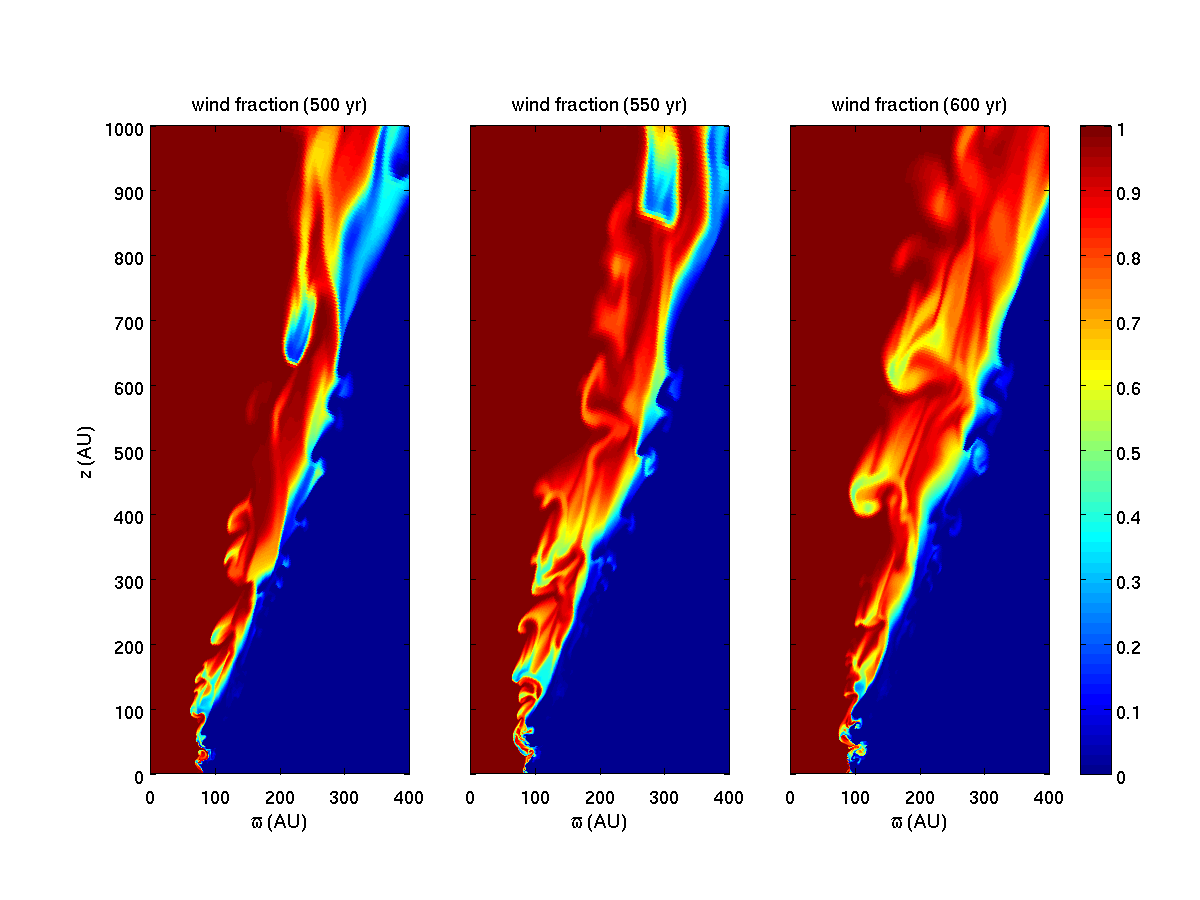}
  \caption{Snapshots of wind fraction map of case {\it b2} (without
    ambient poloidal magnetic field) at $t=500$, 550, and 600\,yr to
    demonstrate the process of mixing between the wind and the ambient.
    Stochastic structures are found to develop in the inner ($z<100$\,AU)
    dense portion of the toroid during interaction with the wind.
    The blobs are subsequently ejected along the flow. Given the
    strong shear between wind and ambient, Kelvin-Helmholtz
    instability could aid in producing the corrugated interface.
    }
  \label{fig_mixingnobp}
\end{figure}

\begin{figure}     
  \includegraphics[width=\textwidth]{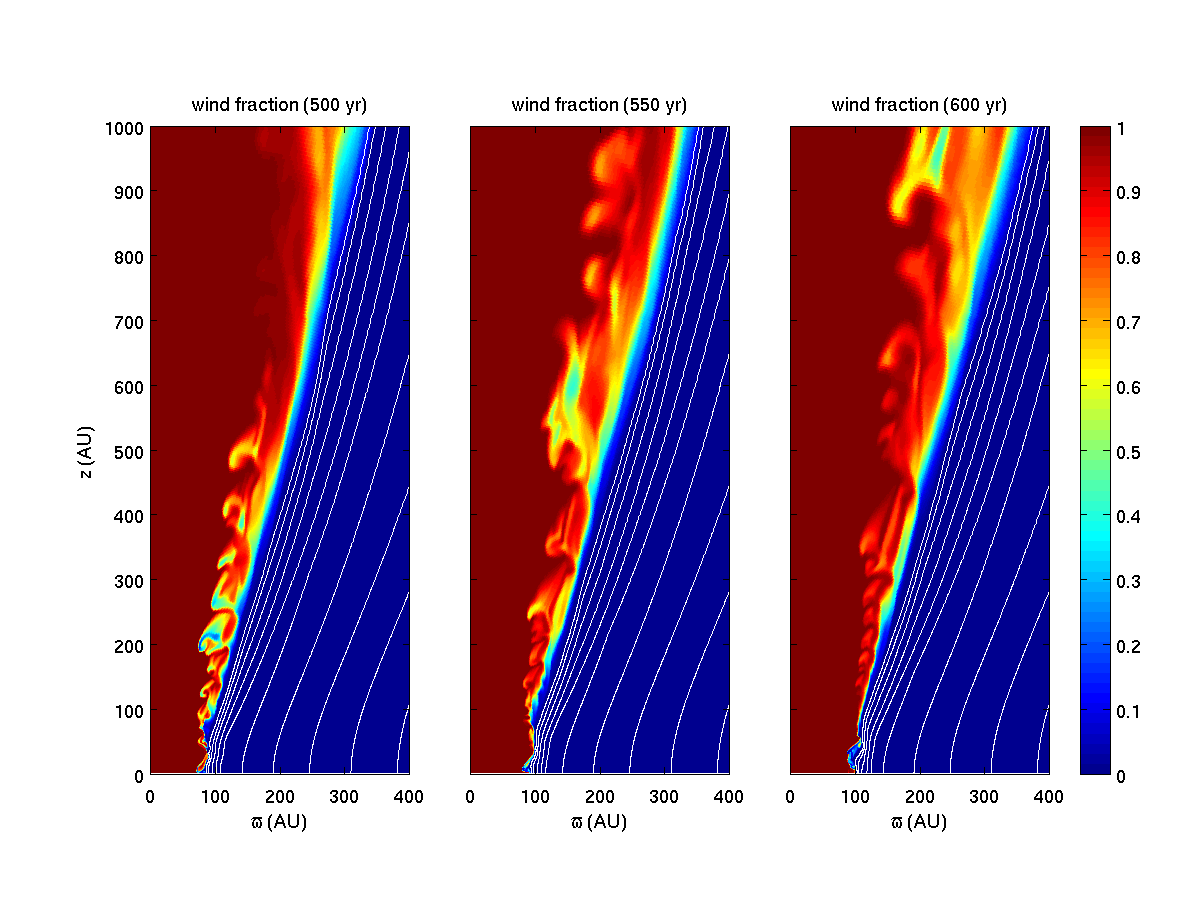}
  \caption{Snapshots of wind fraction map of case {\it b2} (with
    ambient poloidal magnetic field included) at $t=500$, 550, and 600\,yr
    to demonstrate the process of mixing between the wind and the ambient.
    The poloidal field lines are plotted with white lines.
    Stochastic structures are still seen to develop in the inner 
    dense portion of the toroid, but appear to be less significant.
    With the presence of poloidal magnetic field in the toroid, the
    interface is more stable and less corrugated compared with
    Figure \ref{fig_mixingnobp}. 
    }
  \label{fig_mixingbp}
\end{figure}

For a more detailed picture we need to look into the mixing region. 
As described in section \ref{sec_results_basic},
mixing happens within two kinds of regions: a thin transition layer between 
the wind and the ambient, and blobs and feather-like substructures.
The mixing substructures are mostly a result of dense toroid material 
entrained and carried by the wind, 
and we illustrate their development in Figure \ref{fig_mixingnobp}
using case {\it b2} (without ambient poloidal field) as an example. 
We plot the wind fraction at the inner 
few hundred AU for three different epochs $t=500$, 550, and 600\,yr. 
Stochastic, corrugated structures are found to evolve at the
dense base of the toroid as well as on the interface.
Blobs of ambient gas, big or small, can be entrained by the wind and 
subsequently carried by the flow.
Given the strong velocity shear at the wind-ambient interface, 
the corrugated boundary is likely related to the 
Kelvin-Helmholtz instability. 
We have also mentioned in section \ref{sec_results_bp} that the 
presence of poloidal field can help resist the corrugation 
of the swept-up shell. In Figure \ref{fig_mixingbp}
we present wind fraction snapshots similar to those of 
Figure \ref{fig_mixingnobp} for the simulations including poloidal field.
The field lines are plotted with white contours. 
A comparison to Figure \ref{fig_mixingnobp} again shows how
corrugation is suppressed at the interface and a much smoother boundary
is found. Note, however, that even in the presence of a poloidal field, 
there are still feather-like mixing structures growing inside the wind region. 
One reason could be that the poloidal field which helps stabilize the shear
instabilities is only present in the toroid and not in the wind. 
Another could be the much higher density of the toroids as compared 
with the wind.
Any wind material running into the toroids will not go too far
before losing its momentum to the dense swept-up shell. 
The wind boundary therefore appears largely well defined despite the 
complexity of the substructures.

Various types of models have been proposed to explain bipolar molecular outflows, 
and among which wind-driven and jet entrainment are two popular mechanisms 
\citep{cabrit1997}. In the jet-driven scenario, one key question is how 
a collimated jet can drive a less collimated molecular outflow.
Since a turbulent mixing layer between the jet and the ambient is likely 
thin \citep[e.g.,][]{canto1991}, entrainment through a 
jet bowshock \citep[e.g.,][]{raga1993} appears to be more promising. 
Here we would like to note that the setup of our wind simulation is sufficiently different from the jet bowshock scenario so 
that bowshock entrainment is irrelevant here.
In the jet bowshock entrainment model, the molecular outflow is 
identified with the environmental gas entrained into the wake of the bowshock, whereas 
in the wind scenario ambient material is swept up by the primary wind.
There is no prominent bowshock in our wind simulations, which is partly 
due to the lack of ambient material in the polar region of the toroid solution. 
With the tracer field, we have also seen that the mixed material 
within the wind region mainly results from entrainment and/or ejection 
at the wind boundary, especially near the inner dense toroid.
We do not find significant ambient material being incorporated or entrained 
near the head (the polar region) of the outflow. 
In our wind simulations, the wide opening angle of the shell is a 
natural consequence of the wide-angle wind and the toroidal distribution 
of the ambient density, and has little to do 
with the entrainment process going on at the sheared interface.

With improved resolution, in this work we have resolved corrugated 
swept-up shells as well as small blobs of mixed material in 
the wind region that were not seen in paper I. 
The swept-up shells are much smoother in the simulations including poloidal 
field, which is related to the stabilization of 
shear instability by the magnetic field.
To check how an even higher resolution can affect the results, 
we have run simulations for case {\it b2} at a resolution of  
3200$\times$990. 
In the upper panel of Figure \ref{fig_resolution} we present 
the density map at $t=1000$\,yr for the run
without ambient poloidal field. A close comparison to its ordinary 
resolution counterpart (see the upper left panel
of Figure \ref{fig_bpandshell}) shows that the details of the mixing structures 
are quite different in the two runs. 
More complex and fragmented mixing structures are generated in the higher resolution 
simulation, which indicates that the process producing 
these structures is not completely resolved at the current resolution \citep[see, e.g.,][]{frank1996}.
Despite the different fine structures, the overall outflow shape 
and the jet structure in the two simulations are in good agreement.
In the lower panel of Figure \ref{fig_resolution} we present 
the same maps for the run with ambient poloidal field. 
The smoother and thicker shell seen here is almost identical
to its ordinary resolution counterpart (see the lower left panel of 
Figure \ref{fig_bpandshell}), and similar feather-like structures are also
present near the wind boundary. There is also no sign of intense 
stochastic behavior like that seen in the upper panel of
Figure \ref{fig_resolution}. This suggests that the stabilization 
of the wind-ambient interface by the poloidal field is quite robust, 
and this conclusion holds well at least for the resolution we have used.
Finally, it should be noted that the wind in the current simulations 
fans out in all radial directions even at the equator where 
an accreting inner envelope or disk is instead expected. 
Whether this setup could have anything to do 
with the growth of stochastic mixing structures near the inner toroid 
will need to be resolved in future simulations with more realistic 
setup of boundary conditions.

\section{CONCLUSION}\label{sec_conclusion}

\begin{figure}     
  \includegraphics[width=\textwidth]{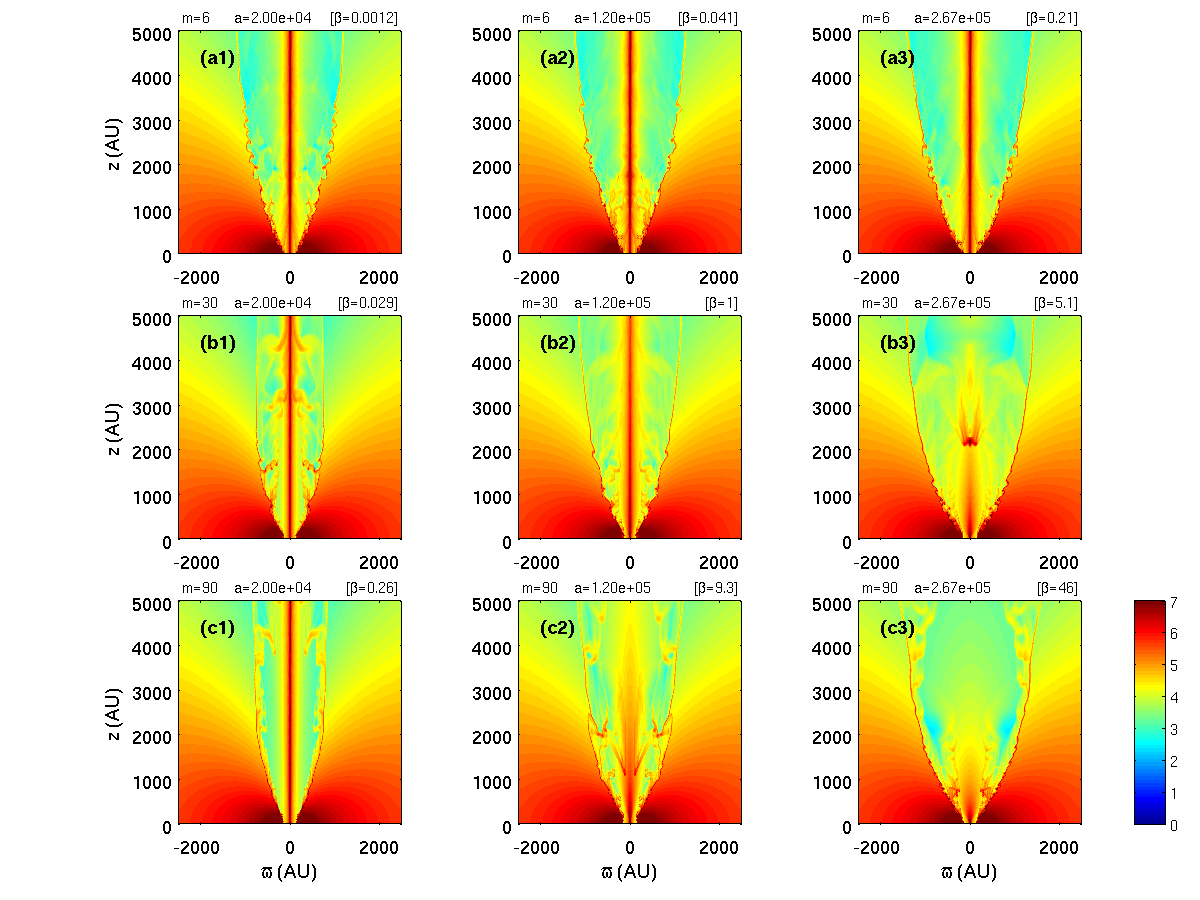}
  \caption{A three-by-three mosaic of density snapshots at $t=1000$\,yr
    for nine different cases. The wind temperature increases from
    $\sim10$\,K in column 1 to $\sim400$\,K and $\sim2000$\,K in
    column 2 and 3, corresponding to $a_{\rm wind}=2\times10^4,
    1.2\times10^5$, and $2.67\times10^5$\,cm\,s$^{-1}$,
    respectively. The strength of the toroidal magnetic field in the
    wind decreases form the strongest $M_{A}=6$ on the top row ($a$)
    to $M_{A}=30$ and 90 in the second and the third row ({\it b} and
    {\it c}). The results demonstrate the effect of varying magnetic
    field and wind temperature in the wind. The jet structure is 
    well-maintained in the magnetic-dominated regime, and flattens 
    with distance in magnetic-dominated regime.}
  \label{fig_mcs1nobp}
\end{figure}

\begin{figure}     
  \includegraphics[width=\textwidth]{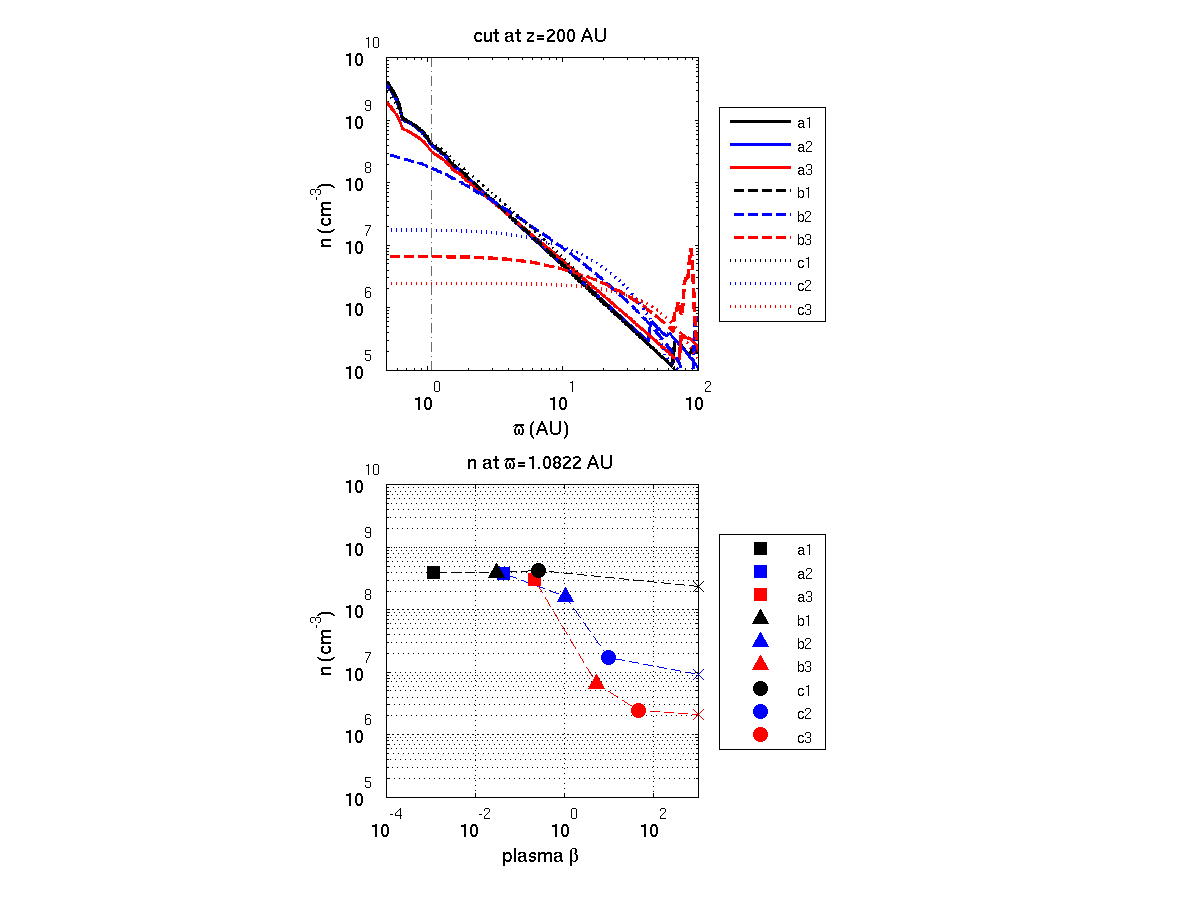}
  \caption{ \textit{Top}: density profile cut perpendicular to the
    outflow axis at $z=200$\,AU for the nine cases
    presented in Figure 2. The modification of the cylindrically
    stratified density profile by thermal pressure is clearly
    seen. Note that the lines for cases {\it a1, b1, c1, a2}, and {\it a3} 
    are very close or even overlap with each other in this plot.   
    \textit{Bottom}: wind density sampled at $\varpi\simeq1$ AU
    plotted against the $\beta_{\rm plasma}$ value for each
    case. Data points with the same wind temperature are connected
    using dashed lines. We also include results of non-magnetized wind 
    simulations at the three temperatures explored (the three cross symbols 
    on the right boundary of the plot), which can be thought of as 
    $\beta_{\rm plasma}=\infty$. 
    In magnetic-dominated regime, the initial density
    is well maintained by the magnetic collimation. In thermal-dominated
    regime, the density approaches what one would
    obtain in a non-magnetized wind.    
    }
  \label{fig_comparedensitydrop}
\end{figure}

\begin{figure}     
  \includegraphics[width=\textwidth]{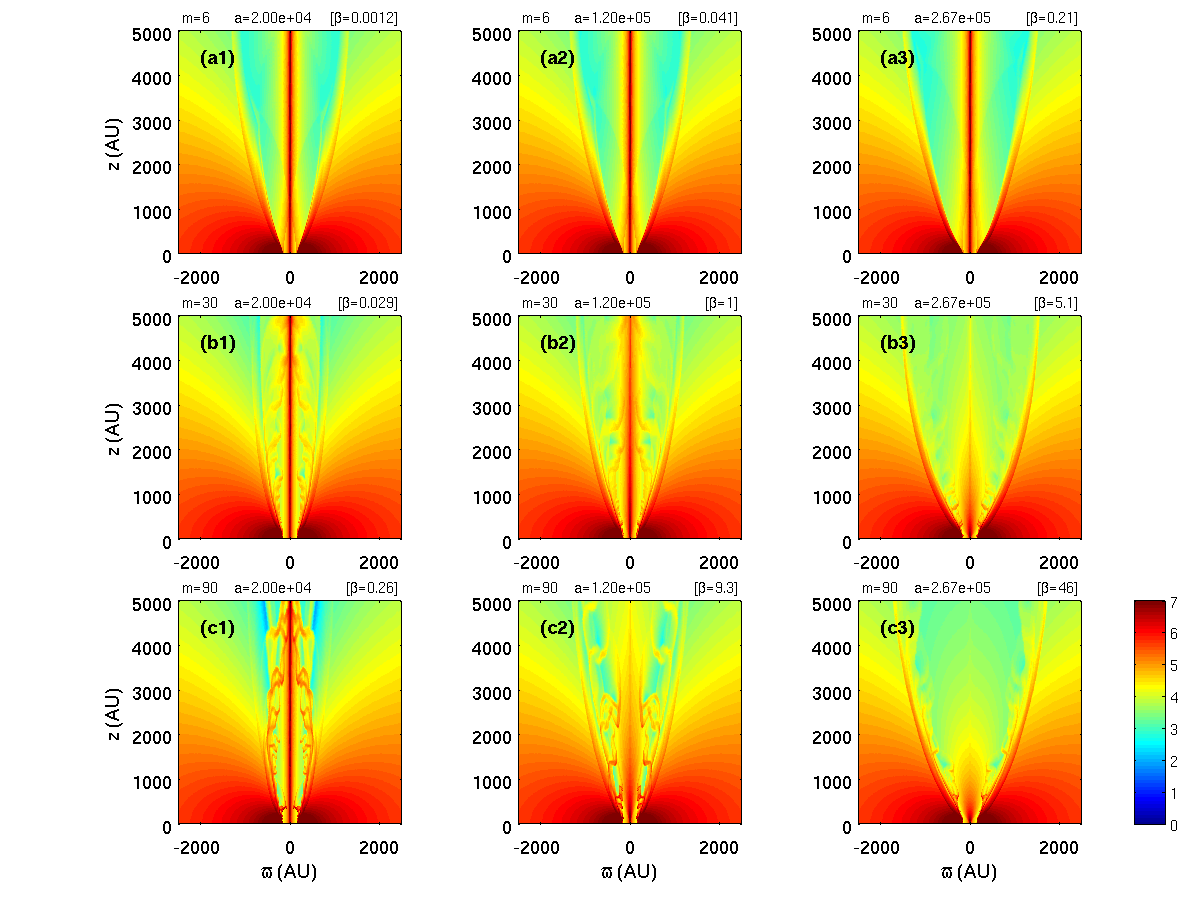}
  \caption{A three-by-three mosaic of density snapshots at $t=1000$\,yr 
  	for nine simulations with ambient poloidal magnetic
    fields. The same combinations of $a_{\rm wind}$ and $M_A$ as in
    Figure \ref{fig_mcs1nobp} are used. The swept-up shell is clearly
    affected by the presence of an ambient poloidal field.}
  \label{fig_mcs1bp}
\end{figure}

\begin{figure}     
  \includegraphics[width=\textwidth]{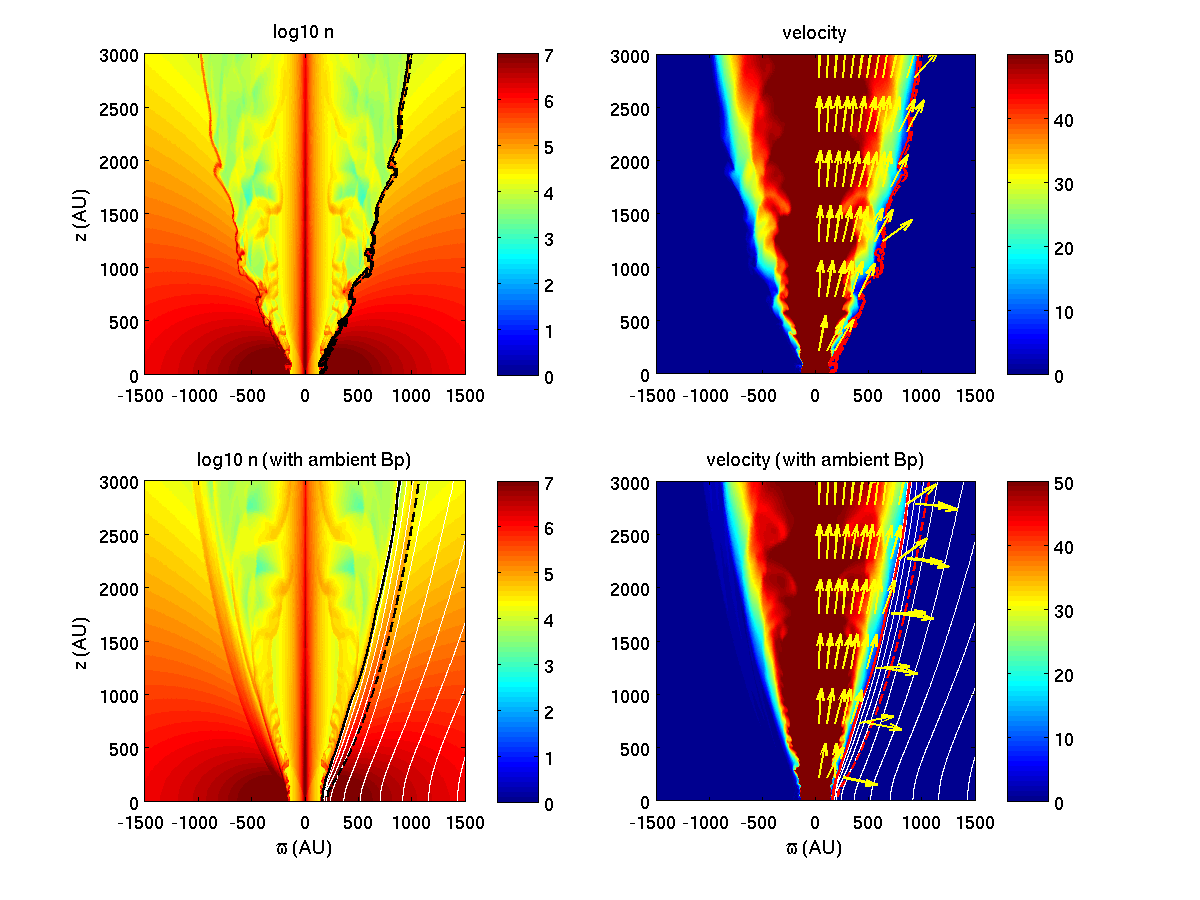}  
  \caption{Density and velocity snapshots at $t=1000$\,yr comparing 
  	simulations ignoring (\textit{upper}) and including (\textit{lower})
  	the ambient poloidal fields. In this example the Alfv\'{e}n Mach
    number $M_{A}$ is 30 and the wind sound speed $a_{\rm wind}$ is
    $1.2\times10^5$\,cm\,s$^{-1}$ ($\sim400$\,K). The thick solid lines
    are contours of 1\% wind fraction, and the poloidal magnetic fields
    are plotted with thin white contours in the lower panels. The
    thick dashed lines are velocity contours of
    $|v_{\rm poloidal}|=2\times10^4$\,cm\,s$^{-1}$, which is the
    ambient sound speed $a_{\rm ambient}$. The flow directions are
    indicated by arrows (only showing those with velocity greater than
    $a_{\rm ambient}$). The modification of shell structures by the 
    presence of poloidal magnetic field is clearly seen. }
  \label{fig_bpandshell}
\end{figure}

\begin{figure}     
	\includegraphics[width=\textwidth]{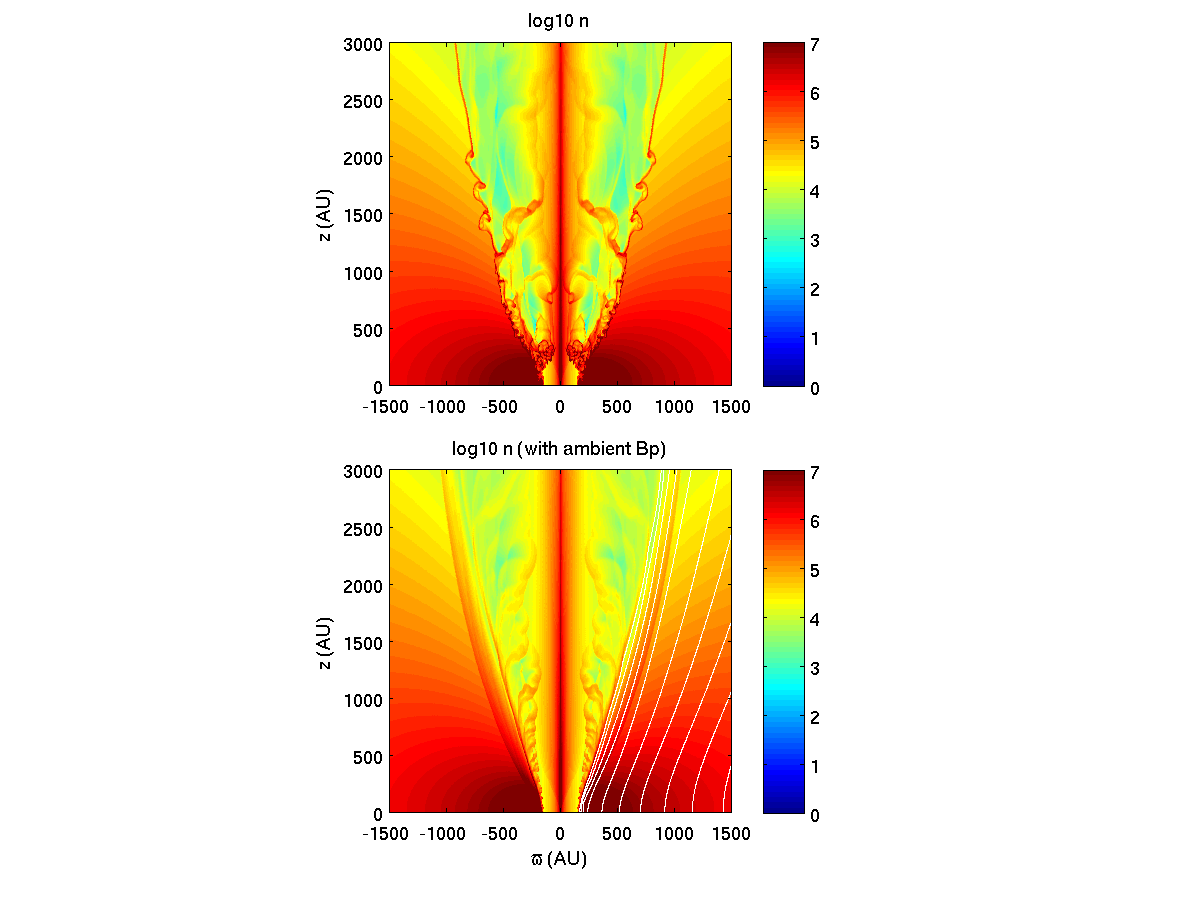}	
	\caption{Density maps of high resolution simulations for case {\it b2} 
		at $t=1000$\,yr. The resolution is 3200$\times$990, which is
		higher than the ordinary 1600$\times$720 
		used throughout this work. The upper and lower panels show the run without 
		and with ambient poloidal field, respectively. 
		The poloidal magnetic field is plotted with thin white lines 
		in the lower panel.
		In the absence of poloidal field, details of the stochastic 
		structures appear quite different from its ordinary resolution counterpart 
		(upper left panel of Figure \ref{fig_bpandshell}).
		The run with ambient poloidal magnetic field, on the other hand, 
		appears quite similar to its ordinary resolution counterpart 
		(lower left panel of Figure \ref{fig_bpandshell}).
	}
	\label{fig_resolution}
\end{figure}

In this work, we aim to understand the relation between outflow
characteristics and their underlying physical factors using MHD
numerical simulations in context of the unified wind model
\citep{shang2006}. In the wind model, the jet and shell
characteristics common to young molecular outflows are results of the
intrinsic stratification of wind density profile and the swept-up
materials during the wind-ambient interaction. Their
detailed morphologies and kinematics may contain information
about the underlying physical conditions of the wind. 
We study the consequences of varying temperature and toroidal
magnetic field in the wind using the ZEUS-TW code. A
series of runs are carried out using a tracer field technique to keep
track of the wind material and to enable the two-temperature
prescription adopted. While keeping the ambient toroids unchanged, 
different combinations of wind temperatures (sound speed
$a_{\rm wind}=2\times10^4$, $1.2\times10^5$, $2.67\times10^5$\,cm\,s$^{-1}$)
and wind Alfv\'{e}n Mach numbers ($M_A=6, 30, 90$) have been
investigated. We also examine the role of poloidal magnetic field of the
toroid by comparing runs with and without including the ambient
poloidal fields. Our findings are as follows:

\begin{enumerate}

  \item For a less magnetized wind, the cylindrically stratified density 
profile that forms the central jet component can be modified by the 
thermal pressure at a high wind temperature. 
In that case, as the wind temperature increases, the wind gas pressure can drive material
away from the dense axis so that the density profile will gradually flatten
as the wind propagates. In extreme cases, the jet-like structure is barely
identifiable beyond several thousand AU (e.g., case {\it c3} in 
Figure \ref{fig_mcs1nobp}).
	
  \item The toroidal magnetic field, on the other hand, tends to
maintain the initial density structure of the wind and appears to
oppose the change made by the thermal pressure. For a strongly magnetized 
wind (such as $M_A=6$), it dominates over thermal pressure even at a
temperature as high as $\sim2000$\,K, and the jet structure appears
unaltered as it propagates away from the source (case {\it a3} in Figure 
\ref{fig_mcs1nobp}). 
	
  \item Therefore in a magnetized wind it is the magnetic
field and thermal pressure that jointly dictate the evolution of the
wind, especially the jet-like density profile. 
For a sufficiently magnetized wind, the jet identity is 
well maintained by the magnetic collimation effect. If the field is not
strong enough to oppose the effects of thermal pressure, the density profile
will evolve and become less axially concentrated as the flow propagates. 
Such effect is more significant for a higher wind temperature. 
Arranging the results in terms of the relative importance 
of thermal and the magnetic pressure may help understand the opposing roles 
of the two factors (Figure \ref{fig_comparedensitydrop}). For the intermediate
case {\it b2}, the change in the density profile is mild and the jet-like
structure is still clearly seen up to a distance of at least 10000\,AU.
	
  \item The poloidal magnetic field that threads the toroid is found to
play a role in the wind-ambient interaction. 
When threaded by the field lines, the toroids
become more resistant to corrugation and a smoother wind-ambient interface is
formed. Also, as the wind pushes the toroids aside, field lines bunch up
and the magnetic pressure is enhanced, making the region more
resistant to compression. The swept-up shell appears thicker as  
compared to the thin layer obtained in the absence of poloidal field. 
The cushioning effect of poloidal field thus
leads to a smoother and thicker swept-up shell (Figure \ref{fig_bpandshell}).
	
\end{enumerate}

In short, our explorations suggest that the density contrast of the
jet is closely related to the relative strengths of the magnetic field and
the thermal pressure in the wind. We also find that an ambient poloidal
field could be the key factor to form a smooth and thicker swept-up
shell. The variety of results also demonstrate the potential
capability of the model in accommodating the rich outflow phenomenon
observed. While the basic principals are clearly demonstrated, whether
they can help constrain physical conditions of molecular outflows in
the real world will depend on more detailed predictions of observable
diagnostics. There are several possible directions to pursue in the
future work. For example, it would also be interesting to check
whether non-ideal MHD effects will result in different behaviors,
especially for the wind-ambient interactions.

\acknowledgments
This work was supported by funds from the Theoretical Institute for Advanced 
Research in Astrophysics (TIARA) through the Academia Sinica and from the 
Ministry of Science and Technology of Taiwan by MOST 102-2119-M-001-008-MY3.

\end{CJK*}
\end{document}